\documentclass{article}
\usepackage{graphicx} % Required for inserting images
\usepackage[utf8]{inputenc} % 'cp1252'-Western, 'cp1251'-Cyrillic, etc.
\usepackage[english]{babel} % 'french', 'german', 'spanish', 'danish', etc.
\usepackage{amsmath}
\usepackage{amssymb}
\usepackage{txfonts}
\usepackage{mathdots}
\usepackage[classicReIm]{kpfonts}
\usepackage{graphicx}
\usepackage{setspace}
\title{\textbf{On the equivalence of two spinodal decomposition criteria with a
case study of
Fe${}_{15}$Co${}_{15}$Ni${}_{35}$Cu${}_{35}$
multicomponent alloy}}
\begin{document}

\maketitle
\noindent Hengwei Luan${}^{1,2,3,*}$, You Wu${}^{4}$, Jingyi Kang${}^{4}$, Liufei Huang${}^{5}$${}^{,}$${}^{6}$, J.H. Luan${}^{7}$, Jinfeng Li${}^{5}$, Yang Shao${}^{4}$, Ke-fu Yao${}^{4}$, Jian Lu${}^{1,2,3,*}$
\noindent 

\noindent ${}^{1}$CityU-Shenzhen Futian Research Institute, Shenzhen 518045, China

\noindent ${}^{2}$Centre for Advanced Structural Materials, City University of Hong Kong Shenzhen Research Institute, Greater Bay Joint Division, Shenyang National Laboratory for Materials Science, Shenzhen 518057, China

\noindent ${}^{3}$Department of Mechanical Engineering, City University of Hong Kong, Tat Chee Avenue, Hong Kong 999077, China

\noindent ${}^{4}$School of Materials Science and Engineering, Tsinghua University, Beijing 100084, China

\noindent ${}^{5}$Institute of Materials, China Academy of Engineering Physics, Mianyang, 621908, China

\noindent ${}^{6}$School of Mechanical Engineering, Xinjiang University, Urumqi, 830017, China

\noindent ${}^{7}$Inter-University 3D Atom Probe Tomography Unit, Department of Mechanical Engineering, City University of Hong Kong, Hong Kong 999077, China.

\noindent 
\\ \hspace*{\fill} \\
\noindent *: Corresponding authors:

\noindent Hengwei Luan: hengluan@um.cityu.edu.hk, luanhengwei6770@163.com;

\noindent Jian Lu: jian.lu@cityu.edu.hk

\noindent 
\eject
\noindent \textbf{Abstract}

\noindent Spinodal decomposition in multicomponent alloys has attracted increasing attention due to its beneficial effect on their mechanical and functional properties and potential applications. Both based on the Cahn-Hillard equation, the reference element method (REM) and the projection matrix method (PMM) are the two main methods to predict the occurrence of spinodal decomposition in multicomponent alloys. In this work, it is mathematically proven that the two methods are equivalent, and therefore the advanced results based on one method can be applied to the other. Based on these methods, the Fe${}_{15}$Co${}_{15}$Ni${}_{35}$Cu${}_{35}$ multicomponent alloy is designed as a case study. Experimental results confirm the spinodal decomposition in the heat-treated alloy, and its strength and ductility are simultaneously enhanced. This work can be the pavement for further theoretical and experimental studies on the spinodal decomposition in multicomponent alloys.
\\ \hspace*{\fill} \\
\\ \hspace*{\fill} \\
\\ \hspace*{\fill} \\
\noindent \textbf{Keywords}

\noindent Multicomponent alloy; spinodal decomposition; reference element method; projection matrix method; high entropy alloy
\eject
\noindent \textbf{Introduction}\\
\noindent  Multicomponent alloys, including high entropy alloys (HEAs), have recently gained significant and increasing attention due to their novel and vast compositional space and excellent mechanical and functional properties[1-7]. The microstructure is of critical importance to their properties, and spinodal decomposition is found to be a common and effective process in multicomponent alloys to alter their microstructures and thus properties[8-15]. Spinodal decomposition refers to the process in which a homogeneous phase becomes unstable and decomposes to usually two phases with the same crystal structure but distinct compositions. Such a process is usually described by the Cahn-Hilliard equation, and the criterion of the occurrence of the spinodal decomposition can be obtained from this equation[16-20]. Although the occurrence of the spinodal decomposition may also be predicted or explained by the phase-field[21-28], molecular dynamic[29-33] or \textit{ab initio}[34-36] methods, the spinodal decomposition criterion derived from the Cahn-Hilliard equation is still helpful since the cost for its calculation is much lower than other methods, and the required input information is easily and readily available, which makes the high-throughput calculation on the novel and vast compositional space of the multicomponent alloys possible[11]. 

\noindent A critical part of the criterion is that the spinodal decomposition criterion must consider the sum of the mole composition fraction of all the components is always equal to one, which limits the possible compositional directions. A simple and widely applied method to realize it is by substituting the composition fraction of one element (denoted as the reference element hereafter) by one minus the sum of the composition fractions of all other elements. For most properties, this treatment can give results independent of the selection of the reference element[37]. However, it is found that such treatment is not rigorous since the absolute value of the calculated driving force of the spinodal decomposition would be dependent on the selection of the reference element[37]. Recently, a more advanced method (denoted as the Reference Element Method (REM) hereafter) has been introduced, where the compositional change is transformed from the Gibbs space into the Cartesian space during the analysis, and it can produce results independent of the selection of the reference element[37-41]. Another method, denoted as the Projection Matrix Method (PMM), can also avoid this "reference element problem"[42-48]. In PMM, a projection matrix is applied to the driving force of the spinodal decomposition to eliminate the component of the driving force that would drive the composition to violate the conservation of mass. Therefore, the conservation of mass is always obeyed, and there is no need to select a reference element. Comparing the two methods, the REM is mainly used in the materials science field, and it has been realized in the Pandat${}^{TM}$ 2022 software[38, 49] recently, while the PMM is mainly used in the mathematical physics and applied mathematics fields for its more straightforward form and easier mathematical treatment. The two methods are supposed to give the same results since they are both valid and they describe the same phenomenon. In a recent work, the authors have shown that alloys with 2 to 10 elements would have the same driving force for spinodal decomposition if the decomposition directions were the same[11]. However, such a result is not satisfying, and a rigorous mathematical proof is still lacking. Therefore, the equivalence of the two methods is not guaranteed, and the advanced theoretical or calculation results based on one method can not be directly transferred to the other.

\noindent In this work, the equivalence of the two spinodal decomposition criteria is mathematically proven, as schematically shown in Fig. S1. Then, the 

\noindent Fe${}_{15}$Co${}_{15}$Ni${}_{35}$Cu${}_{35}$ multicomponent alloy is designed based on the criteria and investigated as a case study for illustration.
\\ \hspace*{\fill} \\

\noindent \textbf{Proof of the equivalence of the two spinodal decomposition criteria}

For a multicomponent alloy with $n$ elements, the composition $\boldsymbol{\mathrm{c}}$ shall be denoted as 
\begin{equation}
\boldsymbol{\mathrm{c}}=\left[\begin{array}{cccc} {c_{1} } & {c_{2} } & {\cdots } & {c_{n} } \end{array}\right]^{T}  \label{1)},
\end{equation}
\noindent where $c_{i} $ is the composition fraction of the $i$${}^{th}$ element, and the upper right corner mark ${}^{T} $ indicates the transpose action. For PMM, the Cahn-Hilliard equation is written as
\begin{equation}
    \frac{\partial }{\partial t}\mathbf{c}=\Delta (-{{\varepsilon }^{2}}\Delta \mathbf{c}+f(\mathbf{c})) \quad in \quad \Omega \label{2}
\end{equation}
\begin{equation}
 \nabla \boldsymbol{\mathrm{c}}\cdot \boldsymbol{\mathrm{n}}=0 \quad and \quad  \nabla (\Delta \boldsymbol{\mathrm{c}})\cdot \boldsymbol{\mathrm{n}}=0 \quad on \quad \partial \Omega  \label{3},
\end{equation}
\noindent where $t$ is the time, $f(\boldsymbol{\mathrm{c}})$\textit{ }is a function of the composition related to the driving force of the spinodal decomposition, $\Omega $ is the geometric space of the alloy (or mathematically, the inner space of the alloy), $\mathrm{\partial}$$\Omega$ is the boundary of the alloy, the \textbf{\textit{n}} is the normal vector of the boundary, and $\varepsilon ^{2} $ is a small positive parameter [42]. The Eq. \eqref{2} has been normalized, and the diffusion term is not explicitly expressed. The $f(\boldsymbol{\mathrm{c}})$ is related to the Gibbs free energy of the alloy as[42]
\begin{equation} \label{4} 
f(\boldsymbol{\mathrm{c}})=P\frac{d}{d\boldsymbol{\mathrm{c}}} G(\boldsymbol{\mathrm{c}}) 
\end{equation} 
where $P$ is the projection matrix and $G(\boldsymbol{\mathrm{c}})$ is the Gibbs free energy as a function of composition of the alloy. The $P$ matrix is given as
\begin{equation}  
 P=I_{n\times n} - \frac{1}{n} J_{n\times n} \label{5},\end{equation}
\noindent where $I_{n\times n} $ is an identity matrix with dimension $n\times n$, and $J_{n\times n} $ is an all-ones matrix with dimension $n\times n$. The $P$ matrix is applied to project the driving force of the spinodal decomposition to the space perpendicular to the vector $[1]_{n} =[1,...,1]\in R_{n} $ (orthogonal complement of $\left\{[1]_{n} \right\}$) by substracting the component vector parallel to $[1]_{n} $. Then, it can be deduced that the criterion for the occurrence of the spinodal decomposition is that the smallest eigenvalue of the following matrix $B_{n\times n} $ (denoted as $\lambda _{\min }^{B} $) is negative[17, 39, 42, 50], where $B$ is defined as
\begin{equation}  
 B=P\frac{d^{2} }{d\boldsymbol{\mathrm{c}}^{2} } G(\boldsymbol{\mathrm{c}}) \label{6}.\end{equation}
\noindent The driving force of the spinodal decomposition would be proportional to $-\lambda _{\min }^{B} $, and the compositional direction of the spinodal decomposition would be the eigenvector $\nu _{\min }^{B} $corresponding to the $\lambda _{\min }^{B} $. It should be noted that the $[\frac{d^{2} }{d\boldsymbol{\mathrm{c}}^{2} } G(\boldsymbol{\mathrm{c}})]^{-1} [1]_{n} $ is always an eigenvector of $B$ with no physical meaning, and it has been excluded from the discussion.
\\ \hspace*{\fill} \\
\noindent As for the REM, we shall take the $n$${}^{th}$ element as the reference element, and the Gibbs free energy shall be given as 
\begin{equation}
 G(\boldsymbol{\mathrm{c}})=G(c_{1} ,...,c_{n-1} ,1-\sum _{i=1}^{n-1}c_{i}  ) \label{7}.\end{equation}
\noindent Then, the fluctuations of the composition in the current Gibbs space are converted to the Cartesian space through a matrix $T$, and the compositional direction and the driving force of the spinodal decomposition in the Cartesian space would be given by the eigenvalue $\lambda ^{\widehat{G}} $ and eigenvector $\nu ^{\widehat{G}} $ of the matrix $\widehat{G}$, which is given by$ $
\begin{equation}
 \widehat{G}=T^{T} \frac{d^{2} G(\boldsymbol{\mathrm{c}})}{d\boldsymbol{\mathrm{c}}_{n-1} {}^{2} } T\label{8}.\end{equation}
\noindent where $\boldsymbol{\mathrm{c}}_{n-1} $ denotes the first $n-1$ terms of the composition. Then, the eigenvectors are transformed back to the Gibbs space by 
\begin{equation}
 \nu ^{R} =T\nu ^{\widehat{G}}  \label{9},\end{equation}
\noindent while the eigenvalues are not changed. It should be noted that the $\nu ^{R} $ is the compositional direction of the first $n-1$ elements, and the composition direction $\nu ^{Rn} $ with $n$ elements should be 
\begin{equation}
 \nu ^{Rn} =\left[\begin{array}{cc} {I_{n-1\times n-1} } & {[0]_{n-1} } \\ {[-1]_{n-1}^{T} } & {1} \end{array}\right]\left[\begin{array}{c} {\nu ^{R} } \\ {0} \end{array}\right]=\left[\begin{array}{c} {\nu ^{R} } \\ {-\sum _{i=1}^{n-1}\nu _{i}^{R}  } \end{array}\right] \label{10}.\end{equation}
\noindent Similar to the PMM method, the criterion for the occurrence of the spinodal decomposition would be the smallest eigenvalue of $\lambda ^{\widehat{G}} $ being negative, and the compositional direction of the spinodal decomposition would be the transformed eigenvector corresponding to that eigenvalue. 
\noindent Previous numerical results indicate that the minimal eigenvalues of the two methods have $\lambda _{\min }^{Rn} =2\lambda _{\min }^{B} $, while the corresponding eigenvectors are identical ($\nu _{\min }^{Rn} =\nu _{\min }^{B} $)[11]. To prove the equivalence of the two methods, we shall prove that such a relationship is valid for all eigenvalues and eigenvectors, which would be a natural result if we can prove that the matrixes of the two methods are equivalent after some transformation.
\noindent First, we shall start from the side of REM. Taking the Eq. \eqref{7} into the $\frac{d^{2} G(\boldsymbol{\mathrm{c}})}{d\boldsymbol{\mathrm{c}}_{n-1} {}^{2} } $ in Eq. \eqref{8}, and by the chain rule, we have
 \begin{equation} \frac{d^{2} G(\boldsymbol{\mathrm{c}})}{dc_{{\rm i}} dc_{j} } =\frac{\partial ^{2} G(\boldsymbol{\mathrm{c}})}{\partial c_{{\rm i}} \partial c_{j} } -\frac{\partial ^{2} G(\boldsymbol{\mathrm{c}})}{\partial c_{j} \partial c_{n} } -\frac{\partial ^{2} G(\boldsymbol{\mathrm{c}})}{\partial c_{{\rm i}} \partial c_{n} } +\frac{\partial ^{2} G(\boldsymbol{\mathrm{c}})}{\partial c_{n}^{2} }  \label{11}.\end{equation}
\noindent For simplicity, we shall divide the $\frac{d^{2} G(\boldsymbol{\mathrm{c}})}{d\boldsymbol{\mathrm{c}}^{2} } $ matrix into 4 parts as
 \begin{equation} \frac{d^{2} G(\boldsymbol{\mathrm{c}})}{d\boldsymbol{\mathrm{c}}^{2} } =\left[\begin{array}{cc} {[\frac{d^{2} G(\boldsymbol{\mathrm{c}})}{dc_{i} dc_{j} } ]_{i,j=1,...,n-1} } & {[\frac{d^{2} G(\boldsymbol{\mathrm{c}})}{dc_{i} dc_{n} } ]_{i=1,...,n-1} } \\ {[\frac{d^{2} G(\boldsymbol{\mathrm{c}})}{dc_{j} dc_{n} } ]_{j=1,...,n-1} } & {\frac{d^{2} G(\boldsymbol{\mathrm{c}})}{dc_{n}^{2} } } \end{array}\right]=\left[\begin{array}{cc} {\ddot{G}_{A} } & {\ddot{G}_{C} } \\ {\ddot{G}_{C}^{T} } & {\ddot{G}_{n} } \end{array}\right] \label{12},\end{equation}
\noindent where the symmetry of the $\frac{d^{2} G(\boldsymbol{\mathrm{c}})}{d\boldsymbol{\mathrm{c}}^{2} } $ matrix is used. Then, together with the Eq. \eqref{11}, the $\frac{d^{2} G(\boldsymbol{\mathrm{c}})}{d\boldsymbol{\mathrm{c}}_{n-1} {}^{2} } $ can be given as
 \begin{equation} \frac{d^{2} G(\boldsymbol{\mathrm{c}})}{d\boldsymbol{\mathrm{c}}_{n-1} {}^{2} } =\ddot{G}_{A} +[-1]_{n-1} \ddot{G}_{C}^{T} +\ddot{G}_{C} [-1]_{n-1}^{T} +\ddot{G}_{n} J_{n-1\times n-1} \label{13}.\end{equation}
\noindent Then, from Eqs. \eqref{8}, \eqref{9} and \eqref{10}, we can get $\widehat{G}\nu ^{\widehat{G}} =\widehat{G}T^{-1} \nu ^{R} =\lambda ^{\widehat{G}} \nu ^{\widehat{G}} =\lambda ^{\widehat{G}} T^{-1} \nu ^{R} $, and therefore 
  \begin{equation}T\widehat{G}T^{-1} \nu ^{R} =\lambda ^{\widehat{G}} \nu ^{R}  \label{14}.\end{equation}
\noindent The Eqs.\eqref{8} and \eqref{13} can be taken into Eq.\eqref{14}, and the result will be compared to the matrix in the side of PMM later.
\noindent On the PMM side, we shall have the following equation from Eqs. \eqref{6} and \eqref{10}
 \begin{equation}\left[\begin{array}{cc} {I_{n-1\times n-1} } & {0} \\ {[1]_{n-1}^{T} } & {1} \end{array}\right]P\left[\begin{array}{cc} {\ddot{G}_{A} } & {\ddot{G}_{C} } \\ {\ddot{G}_{C}^{T} } & {\ddot{G}_{n} } \end{array}\right]\left[\begin{array}{cc} {I_{n-1\times n-1} } & {0} \\ {[-1]_{n-1}^{T} } & {1} \end{array}\right]\left[\begin{array}{c} {\left[\nu ^{B} \right]_{n-1} } \\ {0} \end{array}\right]=\lambda ^{B} \left[\begin{array}{c} {\left[\nu ^{B} \right]_{n-1} } \\ {0} \end{array}\right] \label{15},\end{equation}
\noindent where $\left[\nu ^{B} \right]_{n-1} $ denotes the first $n-1$ terms of $\nu ^{B} $ and the \\
\noindent $\left[\begin{array}{cc} {I_{n-1\times n-1} } & {[0]_{n-1} } \\ {[1]_{n-1}^{T} } & {1} \end{array}\right]\left[\begin{array}{cc} {I_{n-1\times n-1} } & {[0]_{n-1} } \\ {[-1]_{n-1}^{T} } & {1} \end{array}\right]=I_{n\times n} $ is applied. By taking the Eq. \eqref{5} into Eq. \eqref{15}, we can have
 \begin{equation}\begin{aligned}
((I_{n-1\times n-1} -\frac{1}{n} J_{n-1\times n-1} )(\ddot{G}_{A} +\ddot{G}_{C} [-1]_{n-1}^{T} )-\frac{1}{n} [1]_{n-1} (\ddot{G}_{C}^{T} +\ddot{G}_{n} [-1]_{n-1}^{T} ))\left[\nu ^{B} \right]_{n-1} \\=\lambda ^{B} \left[\nu ^{B} \right]_{n-1}  \end{aligned} \label{16}.
\end{equation}
\noindent The equivalence of the two methods requires $\{ \nu ^{R} \} =\{ \left[\nu ^{B} \right]_{n-1} \} $ and $\{ \lambda ^{\widehat{G}} \} =\{ 2\lambda ^{B} \} $, and by comparing the Eq. \eqref{16} with Eq. \eqref{14}, this can be obtained if 
 \begin{equation}(I_{n-1\times n-1} -\frac{1}{n} J_{n-1\times n-1} )(\ddot{G}_{A} +\ddot{G}_{C} [-1]_{n-1}^{T} )-\frac{1}{n} [1]_{n-1} (\ddot{G}_{C}^{T} +\ddot{G}_{n} [-1]_{n-1}^{T} )=\frac{1}{2} T\widehat{G}T^{-1}  \label{17}.\end{equation}
\noindent Taking Eqs. \eqref{8} and \eqref{13} into Eq. \eqref{17} with the lemma (see Supplementary Note 1) that 
 \begin{equation}TT^{T} =2(I_{n-1\times n-1} -\frac{1}{n} J_{n-1\times n-1} ) \label{18)},\end{equation}
\noindent it can be found that both sides of the Eq. \eqref{17} equation equal to
 \begin{equation}\begin{aligned} (I_{n-1\times n-1} -\frac{1}{n} J_{n-1\times n-1} )\ddot{G}_{A} +(I_{n-1\times n-1} -\frac{1}{n} J_{n-1\times n-1} )\ddot{G}_{C} [-1]_{n-1}^{T} \\-\frac{1}{n} [1]_{n-1} \ddot{G}_{C}^{T} +\frac{1}{n} J_{n-1\times n-1} \ddot{G}_{n}   \end{aligned} \label{19)},\end{equation}
\noindent and therefore the equivalence of two spinodal decomposition criteria is proven.
\noindent The equivalence of the two methods broadens the possible choice of methods for material scientists and mathematicians, and the conclusions based on one method can now be applied to the other method directly. For example, based on the PMM method, it has been derived that the condition for spinodal decomposition to produce 3 phases is $\lambda _{\min }^{B} \approx \lambda _{\sec ond{\rm \; }smallest}^{B} <0$[42], and this result has been used to explain why spinodal decomposition into 3 phases is rare[11]. With this proof of the equivalence of the two methods, this condition can be tested by the smallest and the second smallest eigenvalues obtained from the REM method directly. 
\\ \hspace*{\fill} \\
\noindent \textbf{Application of the two criteria to the Fe${}_{15}$Co${}_{15}$Ni${}_{35}$Cu${}_{35}$ multicomponent alloy}

\noindent To numerically and experimentally illustrate the criteria, the Fe${}_{15}$Co${}_{15}$Ni${}_{35}$Cu${}_{35}$ multicomponent alloy is designed as an example. To apply the two criteria to actual alloys, we shall obtain the expression of the $G(\boldsymbol{\mathrm{c}})$ term first. Taking pure elements as references, the $G(\boldsymbol{\mathrm{c}})$ can be given by the regular solution model as
 \begin{equation} G(\boldsymbol{\mathrm{c}})=H(\boldsymbol{\mathrm{c}})-T_{Temp} S(\boldsymbol{\mathrm{c}}) \label{20}, \end{equation} where
 \begin{equation}H(\boldsymbol{\mathrm{c}})=\sum _{i=1}^{n}\sum _{j=1}^{n}2H^{i-j}  c_{i} c_{j}   ,\label{21}\end{equation}
 \begin{equation}S(\boldsymbol{\mathrm{c}})=-R\sum _{i=1}^{n}c_{i} \ln (c_{i} )  ,\label{22} \end{equation}
\noindent $H(\boldsymbol{\mathrm{c}})$is the mixing enthalpy, $S(\boldsymbol{\mathrm{c}})$ is the mixing entropy, $T_{Temp} $ is the temperature, $H^{i-j} $ is the mixing enthalpy per mole between the $i$${}^{th}$ element and $j$${}^{th}$ element (note $H^{i-j} =0$ if $i=j$), and $R$ is the ideal gas constant. Although the regular solution model has been widely used in HEAs [51-53] for its simple form and high data availability, the model has also been criticized, and it may not be widely effective[54]. It should be noted that our proof does not depend on the exact expression of $G(\boldsymbol{\mathrm{c}})$, and extra terms, such as the magnetism term, and other expressions or modifications of $H(\boldsymbol{\mathrm{c}})$ and $S(\boldsymbol{\mathrm{c}})$ can be applied[55]. The values of the $H^{i-j} $ can be given by the Miedema's model[55], and the $H^{i-j} $ between the elements in the Fe${}_{15}$Co${}_{15}$Ni${}_{35}$Cu${}_{35}$ multicomponent alloy is shown in Fig. S2, where the Cu element shows a positive mixing enthalpy with other elements, which may lead to spinodal decomposition.
\noindent With the expression of the $G(\boldsymbol{\mathrm{c}})$ term, the two criteria can be applied to the Fe${}_{15}$Co${}_{15}$Ni${}_{35}$Cu${}_{35}$ multicomponent alloy. As for the PMM, based on the Eqs.\eqref{20}, \eqref{21} and \eqref{22}, the $\frac{d^{2} }{d\boldsymbol{\mathrm{c}}^{2} } G(\boldsymbol{\mathrm{c}})$ term can be given as
 \begin{equation} \frac{{{d}^{2}}}{d{{\mathbf{c}}^{2}}}G(\mathbf{c})=\frac{{{d}^{2}}}{d{{\mathbf{c}}^{2}}}H(\mathbf{c})-{{T}_{Temp}}\frac{{{d}^{2}}}{d{{\mathbf{c}}^{2}}}S(\mathbf{c}) \label{23},\end{equation}
\noindent where 1, 2, 3, and 4 denote Fe, Co, Ni, and Cu elements, respectively. The projection matrix $P$ from Eq.\eqref{5} is 
 \begin{equation} P=I_{4\times 4} - \frac{1}{4} J_{4\times 4}  \label{24}.\end{equation}
\noindent Taking Eqs. \eqref{23} and \eqref{24} into Eq.\eqref{6}, and taking the $T_{Temp} =1100K$ as an example, the $B$ matrix can be calculated as shown in Supplementary Note 2, and the eigenvalues and the corresponding eigenvectors of the $B$ matrix are calculated as
 \begin{equation} \left\{\lambda ^{B} \right\}=\left\{-19984.68,{\rm \; }41778.32,{\rm \; }67854.93\right\} \label{25} \end{equation} and
 \begin{equation} \begin{array}{l} {\left\{\nu ^{B} \right\}={\rm \{ [0.470,\; 0.195,\; 0.178,\; -0.843],\; }} \\ {{\rm [-0.224,\; -0.523,\; 0.819,\; -0.072]}} \\ {{\rm [0.692,\; -0.662,\; -0.217,\; 0.187]\} }} \end{array} \label{26}.\end{equation}
\noindent The negative $\lambda _{\min }^{B} $ indicates the spinodal decomposition would occur, and the corresponding eigenvector indicates the spinodal decomposition would produce a Cu-rich phase and a Fe, Co, and Ni-rich phase, as schematically illustrated in Fig. 1a. The second smallest eigenvalue $\lambda _{\sec ond{\rm \; }smallest}^{B} $ is positive, which indicates that the spinodal decomposition would produce two phases like most other multicomponent alloys[11, 42]. The $\lambda _{\min }^{B} $ and the $\lambda _{\sec ond{\rm \; }smallest}^{B} $ as a function of temperature are plotted in Fig. 1b, which indicates a possible spinodal decomposition start temperature at 1739 K. It should be noted that this temperature is very likely to be higher than the melting point of the alloy, and it actually may indicate a decomposition of the liquid phase. The eigenvector corresponding to $\lambda _{\min }^{B} $, which indicates the compositional direction of the decomposition, is shown in Fig. 1c as a function of temperature. It shows that the two phases are Cu-rich and Fe, Co, and Ni-rich within the whole temperature range.
\noindent As for the REM method, we shall use the same expression of $G(\boldsymbol{\mathrm{c}})$ in Eq.\eqref{20}, and calculate the spinodal decomposition criterion at $T_{Temp} =1100K$. Cu is chosen as the reference element first, and the $\widehat{G}$ matrix can be calculated by Eq.\eqref{8} (see Supplementary Note 2), and the eigenvalues $\{ \lambda ^{\widehat{G}} \} $ and the transformed eigenvectors $\{ \nu ^{R} \} =\{ T\nu ^{\widehat{G}} \} $ are
\begin{equation} \label{27} 
\{ \lambda ^{\widehat{G}} \} =\{ -39969.36{\rm \; }83556.64{\rm \; }135709.86\} 
\end{equation} 
 \begin{equation}\begin{array}{l} {\{ \nu ^{R} \} =\{ [-0.664,-0.275,-0.252],} \\ {[0.316,0.740,-1.158],} \\ {[0.979,-0.936,-0.307]\} } \end{array} \label{28}.\end{equation}
\noindent It can be verified that the values of $\{ \lambda ^{\widehat{G}} \} $ is indeed twice the values of $\left\{\lambda ^{B} \right\}$, and the $\{ \nu ^{R} \} =\{ T\nu ^{\widehat{G}} \} $ is the same as $\left\{\nu ^{B} \right\}$ after adding the Cu component and normalization. The Fe, Co, or Ni element can also be chosen as the reference element, and the results are listed in Supplementary Note 2. It can be seen that the $\{ \lambda ^{\widehat{G}} \} $ and the normalized transformed eigenvectors $\{ \nu ^{R} \} $ are not affected by the selection of the reference element. These results numerically confirm the equivalence of two spinodal decomposition criteria and predict the spinodal decomposition of the Fe${}_{15}$Co${}_{15}$Ni${}_{35}$Cu${}_{35}$ multicomponent alloy at 1100 K.
\\ \hspace*{\fill} \\
\noindent \textbf{Experimental analysis of the Fe${}_{15}$Co${}_{15}$Ni${}_{35}$Cu${}_{35}$ multicomponent alloy}

\noindent To experimentally confirm the predicted spinodal decomposition of the \\Fe${}_{15}$Co${}_{15}$Ni${}_{35}$Cu${}_{35}$ multicomponent alloy, the Fe${}_{15}$Co${}_{15}$Ni${}_{35}$Cu${}_{35}$ multicomponent alloy ingot is prepared by arc-melting and copper mould casting and analyzed (details in Supplementay Note 3). The X-ray diffraction (XRD) result shows that the as-prepared alloy has a single face-centered cubic (FCC) phase (a= 0.3592 nm) microstructure (Fig. 1d), and the single-phase microstructure is supposed to be caused by the fast cooling rate of the copper mould casting that dynamically prevents the spinodal decomposition process. To introduce the spinodal decomposition, the as-cast alloy was heat-treated at 1100 K for 1 hour, and the XRD result shows that the heat-treated alloy has two FCC phases (Fig. 1d) with very close lattice parameters (a= 0.3615 nm and 0.3577 nm), which indicates that the spinodal decomposition happened as predicted. Scanning electron microscope (SEM) images show that these alloys keep a dendrite microstructure with a secondary dendrite arm spacing of $\mathrm{\sim}$4 $\muup$m before and after the heat-treatment (Fig. S3). To observed the spinodal decomposition microstructure, the as-prepared and heat-treated alloys are observed by transmission electron microscopy (TEM) and energy dispersive spectroscopy (EDS) analysis. The heat-treated sample shows a maze-like microstructure with a spacing of $\mathrm{\sim}$40 nm (Fig. 2a), which is not observed in the TEM image of the as-prepared alloy (Fig. S4), and the selected area electron diffraction (SAED) image shows that both phases are FCC phases with indistinguishable lattice parameter and crystallographic orientation in TEM as confirmed by the coherent high-resolution TEM image (Fig. S5). These results show that the heat-treated alloy has a typical spinodal decomposition microstructure as predicted. The EDS line scan (Fig. S6) reveals that the spinodal decomposition produces a Cu-rich phase and a Fe, Co, and Ni-rich phase, which agrees with the prediction of the compositional direction. To better observe the composition fluctuation of the spinodal decomposition microstructure, the atom probe tomography (APT) analysis is applied to the heat-treated sample, as shown in Fig. 2c and d. The APT result clearly shows that the heat-treated sample has a Cu-rich phase and a Fe, Co, and Ni-rich phase with a transition region of $\mathrm{\sim}$7 nm (Fig. 2d). This result confirms the spinodal decomposition microstructure and the predicted compositional direction.

\noindent To show the effect of the spinodal decomposition on the properties, the tensile mechanical property of the as-prepared and heat-treated samples are tested as shown in Fig. 3a. The result shows that the spinodal decomposition slightly increases the yield strength (326.7 MPa to 327.5 MPa), while the ultimate tensile strength (573.0 MPa to 583.7 MPa) and the elongation to failure (16.9 \% to 19.5 \%) are significantly increased. Similar increased strength has also been observed in other alloys with spinodal decomposition[12-14, 32, 56]. The increased strength is confirmed by the increased Vicker's hardness ($178.1\pm6.6$ HV to $181.5\pm4.0$ HV). The strain-hardening rate curve (inset in Fig. 3a) shows that the heat-treated sample has a significantly higher strain-hardening rate at the initial stage of deformation, and it gradually decreases with more strain, while the strain-hardening rate of the as-prepared sample remains stable. The magnetic properties of the as-prepared and heat-treated samples are also analyzed by a vibrating sample magnetometer (VSM) as shown in Fig. 3b. The two samples show close saturation magnetization at $\mathrm{\sim}$795 Am${}^{2}$/kg, while the coercive force of the heat-treated sample ($\mathrm{\sim}$3.4 kA/m) is much higher than the as-prepared sample ($\mathrm{\sim}$0.45 kA/m), which is supposed to be caused by the hindering effect of the Cu-rich phase on the arrangement motion of the magnetic domain wall[57-59]. A more detailed analysis would be beneficial for understanding the effect of spinodal decomposition on the properties, but it would be beyond the focus of this work. In short, we have experimentally confirmed the spinodal decomposition of the Fe${}_{15}$Co${}_{15}$Ni${}_{35}$Cu${}_{35}$ multicomponent alloy.
\\ \hspace*{\fill} \\
\noindent \textbf{Conclusion}\\
\noindent In summary, we have successfully demonstrated the equivalence of the reference element method and the projection matrix method for predicting spinodal decomposition in multicomponent alloys through rigorous mathematical proof. Then, the equivalence of the two methods is numerically verified, and the spinodal decomposition of the Fe${}_{15}$Co${}_{15}$Ni${}_{35}$Cu${}_{35}$ multicomponent alloy is predicted. Experimental results confirm the occurrence of the predicted spinodal decomposition, and the improved mechanical properties are observed. The proof in this work can link the results based on one method to the other, which is helpful for further theoretical and experimental studies, and the two criteria can function as powerful tools for further investigations.\textbf{}

\noindent 

\noindent \eject 

\noindent \includegraphics*[width=4.5in, height=3.5in]{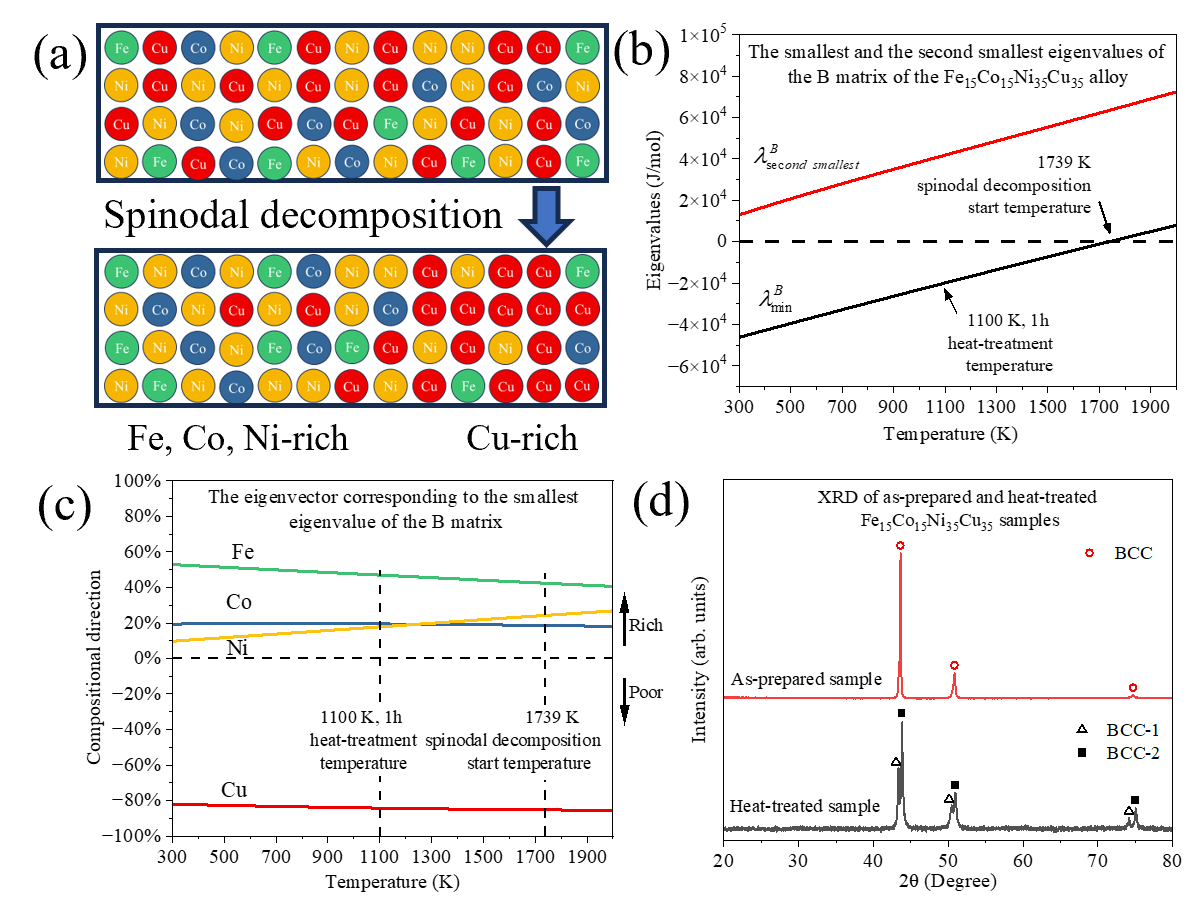}

\noindent \textbf{Fig. 1. Prediction and the XRD results of the Fe${}_{15}$Co${}_{15}$Ni${}_{35}$Cu${}_{35}$ multicomponent alloy. }(a) Schematic diagram of the spinodal decomposition of the Fe${}_{15}$Co${}_{15}$Ni${}_{35}$Cu${}_{35}$ multicomponent alloy. (b) The smallest and the second smallest eigenvalues of the $B$ matrix as a function of temperature. (c) The eigenvector corresponding to the smallest eigenvalue of the $B$ matrix as a function of temperature. (d) The XRD patterns of the as-prepared and heat-treated Fe${}_{15}$Co${}_{15}$Ni${}_{35}$Cu${}_{35}$ multicomponent alloy.

\noindent 

\noindent \eject 

\noindent \includegraphics*[width=4.50in, height=2.3in]{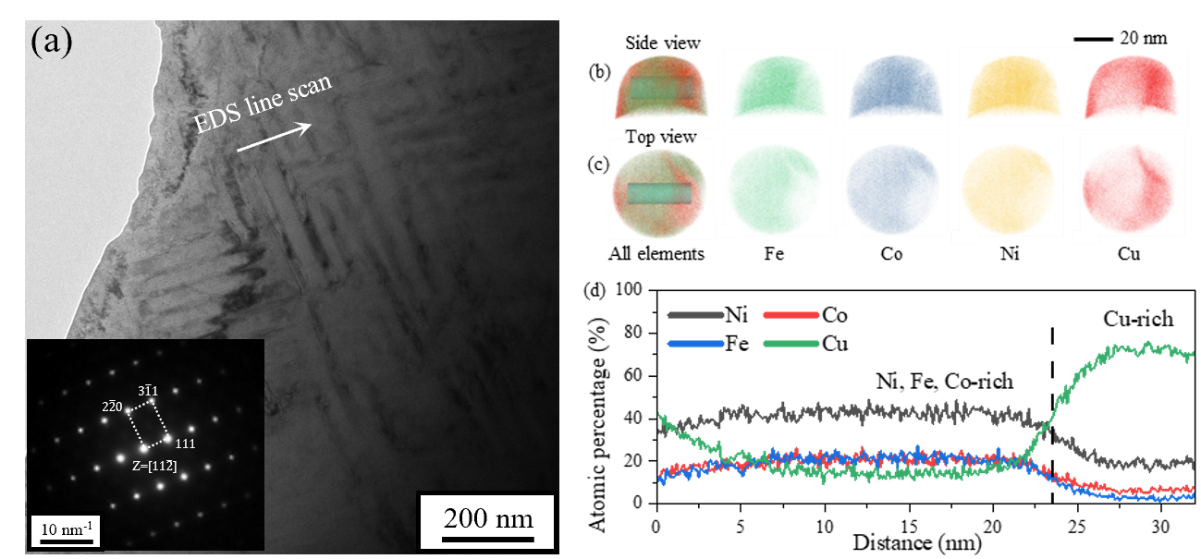}

\noindent \textbf{Fig. 2. TEM and APT analysis of the heat-treated Fe${}_{15}$Co${}_{15}$Ni${}_{35}$Cu${}_{35}$ multicomponent alloy. }(a) Bright-field image of the heat-treated Fe${}_{15}$Co${}_{15}$Ni${}_{35}$Cu${}_{35}$ multicomponent alloy. Inset: The SAED pattern.\textbf{ }(b) Side view of the heat-treated Fe${}_{15}$Co${}_{15}$Ni${}_{35}$Cu${}_{35}$ multicomponent alloy. The analyzed cylinder region of interest is shown in the "all elements" image. (c) Top view of the same sample. (d) The composition of the analyzed cylinder region of interest.

\noindent \eject 

\noindent \includegraphics*[width=4.5in, height=2.0in]{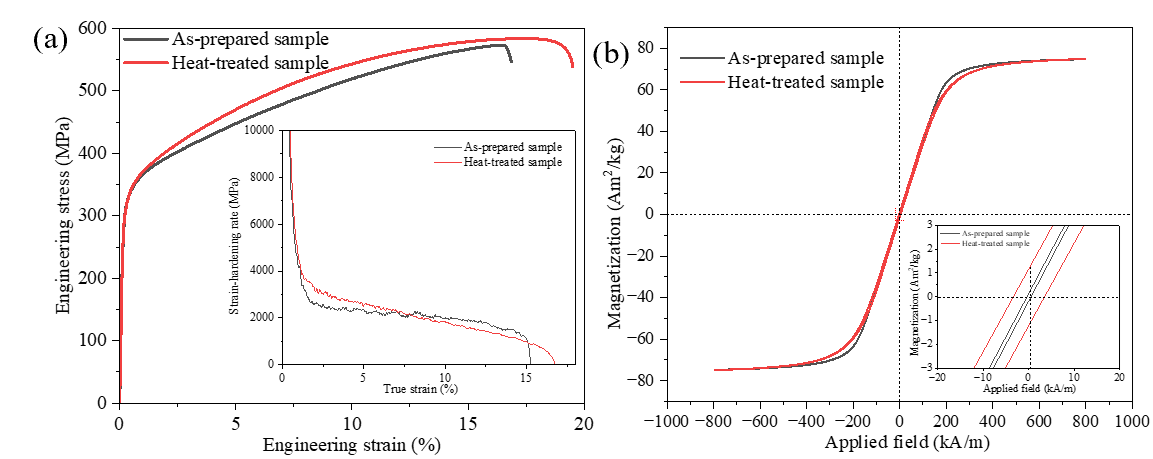}

\noindent \textbf{Fig. 3. Mechanical and magnetic properties of the as-prepared and heat-treated Fe${}_{15}$Co${}_{15}$Ni${}_{35}$Cu${}_{35}$ multicomponent alloy. }(a) Tensile stress-strain curve of the samples. Inset: The strain-hardening rate of the samples. (b) VSM magnetization curves of the samples.

\noindent \eject 
\noindent \textbf{Acknowledgments}

\noindent H.L would like to thank the help of Mr. Zhihao JIANG and an anonymous
master of mathematics. Atom probe tomography research was conducted by
Dr. J.H. Luan at the Inter-University 3D Atom Probe Tomography Unit of
City University of Hong Kong, which is supported by the CityU grant
9360161.

\noindent \eject 
\noindent \textbf{References}

{[}1{]} Y. Yao, Z. Huang, P. Xie, S.D. Lacey, R.J. Jacob, H. Xie, F.
Chen, A. Nie, T. Pu, M. Rehwoldt, D. Yu, M.R. Zachariah, C. Wang, R.
Shahbazian-Yassar, J. Li, L. Hu, Carbothermal shock synthesis of
high-entropy-alloy nanoparticles, Science 359(6383) (2018) 1489-1494.

{[}2{]} B. Gludovatz, A. Hohenwarter, D. Catoor, E.H. Chang, E.P.
George, R.O. Ritchie, A fracture-resistant high-entropy alloy for
cryogenic applications, Science 345(6201) (2014) 1153-8.

{[}3{]} T. Yang, Y.L. Zhao, Y. Tong, Z.B. Jiao, J. Wei, J.X. Cai, X.D.
Han, D. Chen, A. Hu, J.J. Kai, K. Lu, Y. Liu, C.T. Liu, Multicomponent
intermetallic nanoparticles and superb mechanical behaviors of complex
alloys, Science 362(6417) (2018) 933-937.

{[}4{]} Z. Li, K.G. Pradeep, Y. Deng, D. Raabe, C.C. Tasan, Metastable
high-entropy dual-phase alloys overcome the strength-ductility
trade-off, Nature 534(7606) (2016) 227-30.

{[}5{]} Z. Lei, X. Liu, Y. Wu, H. Wang, S. Jiang, S. Wang, X. Hui, Y.
Wu, B. Gault, P. Kontis, D. Raabe, L. Gu, Q. Zhang, H. Chen, H. Wang, J.
Liu, K. An, Q. Zeng, T.G. Nieh, Z. Lu, Enhanced strength and ductility
in a high-entropy alloy via ordered oxygen complexes, Nature 563(7732)
(2018) 546.

{[}6{]} J. Ren, Y. Zhang, D. Zhao, Y. Chen, S. Guan, Y. Liu, L. Liu, S.
Peng, F. Kong, J.D. Poplawsky, G. Gao, T. Voisin, K. An, Y.M. Wang, K.Y.
Xie, T. Zhu, W. Chen, Strong yet ductile nanolamellar high-entropy
alloys by additive manufacturing, Nature 608(7921) (2022) 62-68.

{[}7{]} P. Shi, R. Li, Y. Li, Y. Wen, Y. Zhong, W. Ren, Z. Shen, T.
Zheng, J. Peng, X. Liang, P. Hu, N. Min, Y. Zhang, Y. Ren, K. Liaw
Peter, D. Raabe, Y.-D. Wang, Hierarchical crack buffering triples
ductility in eutectic herringbone high-entropy alloys, Science 373(6557)
(2021) 912-918.

{[}8{]} Y. Zhang, Z. Chen, D.D. Cao, J.Y. Zhang, P. Zhang, Q. Tao, X.Q.
Yang, Concurrence of spinodal decomposition and nano-phase precipitation
in a multi-component AlCoCrCuFeNi high-entropy alloy, Journal of
Materials Research and Technology-Jmr\&T 8(1) (2019) 726-736.

{[}9{]} Z. Rao, B. Dutta, F. Körmann, W. Lu, X. Zhou, C. Liu, A.K.
Silva, U. Wiedwald, M. Spasova, M. Farle, D. Ponge, B. Gault, J.
Neugebauer, D. Raabe, Z. Li, Beyond Solid Solution High‐Entropy Alloys:
Tailoring Magnetic Properties via Spinodal Decomposition, Adv. Funct.
Mater. 31(7) (2020) 2007668.

{[}10{]} K. Kadirvel, S.R. Koneru, Y.Z. Wang, Exploration of spinodal
decomposition in multi-principal element alloys (MPEAs) using CALPHAD
modeling, Scripta Mater. 214 (2022) 114657.

{[}11{]} H. Luan, L. Huang, J. Kang, B. Luo, X. Yang, J. Li, Z. Han, J.
Si, Y. Shao, J. Lu, K.-F. Yao, Spinodal decomposition and the
pseudo-binary decomposition in high-entropy alloys, Acta Mater. 248
(2023) 118775.

{[}12{]} Y. Dong, X.X. Gao, Y.P. Lu, T.M. Wang, T.J. Li, A
multi-component AlCrFe\textsubscript{2}Ni\textsubscript{2} alloy with
excellent mechanical properties, Mater. Lett. 169 (2016) 62-64.

{[}13{]} W.R. Wang, W.L. Wang, S.C. Wang, Y.C. Tsai, C.H. Lai, J.W. Yeh,
Effects of Al addition on the microstructure and mechanical property of
Al\textsubscript{x}CoCrFeNi high-entropy alloys, Intermetallics 26
(2012) 44-51.

{[}14{]} Z.R. Zhang, H. Zhang, Y. Tang, L. Zhu, Y.C. Ye, S. Li, S.X.
Bai, Microstructure, mechanical properties and energetic characteristics
of a novel high-entropy alloy HfZrTiTa\textsubscript{0.53}, Mater. Des.
133 (2017) 435-443.

{[}15{]} T. Borkar, V. Chaudhary, B. Gwalani, D. Choudhuri, C.V. Mikler,
V. Soni, T. Alam, R.V. Ramanujan, R. Banerjee, A Combinatorial Approach
for Assessing the Magnetic Properties of High Entropy Alloys: Role of Cr
in AlCo\textsubscript{x}Cr\textsubscript{1-x}FeNi, Adv. Eng. Mater.
19(8) (2017) 1700048.

{[}16{]} D.E. Laughlin, K. Hono, Physical Metallurgy, Fifth ed.,
Elsevier, Netherland, 2014.

{[}17{]} J.E. Morral, J.W. Cahn, Spinodal Decomposition in Ternary
Systems, Acta Metall. 19(10) (1971) 1037-1045.

{[}18{]} J.W. Cahn, Free Energy of a Nonuniform System. II.
Thermodynamic Basis, The J. Chem. Phys. 30(5) (1959) 1121-1124.

{[}19{]} J.W. Cahn, Phase Separation by Spinodal Decomposition in
Isotropic Systems, The J. Chem. Phys. 42(1) (1965) 93-99.

{[}20{]} J.W. Cahn, J.E. Hilliard, Free Energy of a Nonuniform System.
I. Interfacial Free Energy, The J. Chem. Phys. 28(2) (1958) 258-267.

{[}21{]} J.L. Li, Z. Li, Q. Wang, C. Dong, P.K. Liaw, Phase-field
simulation of coherent BCC/B2 microstructures in high entropy alloys,
Acta Mater. 197 (2020) 10-19.

{[}22{]} D. Fan, L.-Q. Chen, Diffusion-controlled grain growth in
two-phase solids, Acta Mater. 45(8) (1997) 3297-3310.

{[}23{]} N. Moelans, B. Blanpain, P. Wollants, An introduction to
phase-field modeling of microstructure evolution, Calphad 32(2) (2008)
268-294.

{[}24{]} D. Tourret, H. Liu, J. LLorca, Phase-field modeling of
microstructure evolution: Recent applications, perspectives and
challenges, Prog. Mater Sci. 123 (2022) 100810.

{[}25{]} L.-Q. Chen, Phase-field models for microstructure evolution,
Annual review of materials research 32(1) (2002) 113-140.

{[}26{]} L.-q. Chen, Computer simulation of spinodal decomposition in
ternary systems, Acta metallurgica et materialia 42(10) (1994)
3503-3513.

{[}27{]} M. Tang, A. Karma, Surface Modes of Coherent Spinodal
Decomposition, Phys. Rev. Lett. 108(26) (2012) 265701.

{[}28{]} M. Salvalaglio, M. Bouabdellaoui, M. Bollani, A. Benali, L.
Favre, J.-B. Claude, J. Wenger, P. de Anna, F. Intonti, A. Voigt, M.
Abbarchi, Hyperuniform Monocrystalline Structures by Spinodal
Solid-State Dewetting, Phys. Rev. Lett. 125(12) (2020) 126101.

{[}29{]} S.K. Das, S. Puri, J. Horbach, K. Binder, Spinodal
decomposition in thin films: Molecular-dynamics simulations of a binary
Lennard-Jones fluid mixture, Physical Review E 73(3) (2006) 031604.

{[}30{]} P.K. Jaiswal, S. Puri, S.K. Das, Surface-directed spinodal
decomposition: A molecular dynamics study, Physical Review E 85(5)
(2012) 051137.

{[}31{]} S. Toxvaerd, Molecular dynamics simulations of spinodal
decomposition in films of binary mixtures, Phys. Rev. Lett. 83(25)
(1999) 5318.

{[}32{]} T. Xin, Y. Zhao, R. Mahjoub, J. Jiang, A. Yadav, K. Nomoto, R.
Niu, S. Tang, F. Ji, Z. Quadir, D. Miskovic, J. Daniels, W. Xu, X. Liao,
L.-Q. Chen, K. Hagihara, X. Li, S. Ringer, M. Ferry, Ultrahigh specific
strength in a magnesium alloy strengthened by spinodal decomposition,
Science Advances 7(23) (2021) eabf3039.

{[}33{]} Y. He, P. Yi, M.L. Falk, Critical Analysis of an FeP Empirical
Potential Employed to Study the Fracture of Metallic Glasses, Phys. Rev.
Lett. 122(3) (2019) 035501.

{[}34{]} T. Fukushima, K. Sato, H. Katayama-Yoshida, P.H. Dederichs,
\emph{Ab initio} study of spinodal decomposition in (Zn, Cr)Te, physica
status solidi (a) 203(11) (2006) 2751-2755.

{[}35{]} R.F. Zhang, S. Veprek, Phase stabilities and spinodal decomposition in the Cr\textsubscript{1-x}Al\textsubscript{x}N system
studied by ab initio LDA and thermodynamic modeling: Comparison with the
Ti\textsubscript{1-x}Al\textsubscript{x}N and
TiN/Si\textsubscript{3}N\textsubscript{4} systems, Acta Mater. 55(14)
(2007) 4615-4624.

{[}36{]} P.H. Mayrhofer, D. Music, J.M. Schneider, \emph{Ab initio}
calculated binodal and spinodal of cubic
Ti\textsubscript{1-x}Al\textsubscript{x}N, Appl. Phys. Lett. 88(7)
(2006) 071922.

{[}37{]} J.E. Morral, S.L. Chen, Stability of High Entropy Alloys to
Spinodal Decomposition, J Phase Equilibria Diffus 42(5) (2021) 673-695.

{[}38{]} S.R. Koneru, K. Kadirvel, Y. Wang, High-Throughput Design of
Multi-Principal Element Alloys with Spinodal Decomposition Assisted
Microstructures, J Phase Equilibria Diffus 43(6) (2022) 753-763.

{[}39{]} D. De Fontaine, An analysis of clustering and ordering in
multicomponent solid solutions---I. Stability criteria, J. Phys. Chem.
Solids 33(2) (1972) 297-310.

{[}40{]} D. De Fontaine, An analysis of clustering and ordering in
multicomponent solid solutions---II fluctuations and kinetics, J. Phys.
Chem. Solids 34(8) (1973) 1285-1304.

{[}41{]} P. Singh, A.V. Smirnov, D.D. Johnson, Atomic short-range order
and incipient long-range order in high-entropy alloys, Phys. Rev. B
91(22) (2015) 224204.

{[}42{]} S. Maier-Paape, B. Stoth, T. Wanner, Spinodal decomposition for
multicomponent Cahn-Hilliard systems, J Stat Phys 98(3-4) (2000)
871-896.

{[}43{]} C.M. Elliott, S. Luckhaus, A generalised diffusion equation for
phase separation of a multi-component mixture with interfacial free
energy, IMA Preprint Series 887 (1991).

{[}44{]} H. Garcke, On a Cahn--Hilliard model for phase separation with
elastic misfit, Annales de l\textquotesingle Institut Henri Poincaré C,
Analyse non linéaire 22(2) (2005) 165-185.

{[}45{]} J.P. Desi, H.H. Edrees, J.J. Price, E. Sander, T. Wanner, The
Dynamics of Nucleation in Stochastic Cahn--Morral Systems, SIAM Journal
on Applied Dynamical Systems 10(2) (2011) 707-743.

{[}46{]} H. Garcke, On mathematical models for phase separation in
elastically stressed solids, Math. Inst. der Univ., 2000.

{[}47{]} A. Miranville, G. Schimperna, Generalized Cahn-Hilliard
equations for multicomponent alloys, Advances in Mathematical Sciences
and Applications 19(1) (2009) 131.

{[}48{]} H. Garcke, On Cahn---Hilliard systems with elasticity,
Proceedings of the Royal Society of Edinburgh Section A: Mathematics
133(2) (2003) 307-331.

{[}49{]} W. Cao, S.-L. Chen, F. Zhang, K. Wu, Y. Yang, Y. Chang, R.
Schmid-Fetzer, W. Oates, PANDAT software with PanEngine, PanOptimizer
and PanPrecipitation for multi-component phase diagram calculation and
materials property simulation, Calphad 33(2) (2009) 328-342.

{[}50{]} J.E. Morral, Stability Limits for Ternary Regular System, Acta
Metall. 20(8) (1972) 1069-1076.

{[}51{]} A.B. Melnick, V.K. Soolshenko, Thermodynamic design of
high-entropy refractory alloys, J. Alloys Compd. 694 (2017) 223-227.

{[}52{]} Y. Zhang, T.T. Zuo, Z. Tang, M.C. Gao, K.A. Dahmen, P.K. Liaw,
Z.P. Lu, Microstructures and properties of high-entropy alloys, Prog.
Mater Sci. 61 (2014) 1-93.

{[}53{]} B.S. Murty, J.W. Yeh, S. Raganathan, High-Entropy Alloys,
Springer, Boston, 2014.

{[}54{]} D.B. Miracle, O.N. Senkov, A critical review of high entropy
alloys and related concepts, Acta Mater. 122 (2017) 448-511.

{[}55{]} A. Takeuchi, A. Inoue, Mixing enthalpy of liquid phase
calculated by miedema\textquotesingle s scheme and approximated with
sub-regular solution model for assessing forming ability of amorphous
and glassy alloys, Intermetallics 18(9) (2010) 1779-1789.

{[}56{]} Y.-J. Liang, L. Wang, Y. Wen, B. Cheng, Q. Wu, T. Cao, Q. Xiao,
Y. Xue, G. Sha, Y. Wang, Y. Ren, X. Li, L. Wang, F. Wang, H. Cai,
High-content ductile coherent nanoprecipitates achieve ultrastrong
high-entropy alloys, Nat. Commun. 9(1) (2018) 4063.

{[}57{]} D. Jiles, Introduction to magnetism and magnetic materials, CRC
press2015.

{[}58{]} K. Jenkins, M. Lindenmo, Precipitates in electrical steels, J.
Magn. Magn. Mater. 320(20) (2008) 2423-2429.

{[}59{]} L. Shang, H.L. Yang, H. Xu, Y.G. Li, The influence factors of
silicon steel magnetic property, Advanced Materials Research, 2014, pp.
345-348.
\end{document}